# High pressure synthesis and structural study of AuGa$_2$ intermetallic compound

Azkar Saeed Ahmad[a,b,*], Zhuoyang Yu[a,1], Sizhe Wang[a,1], Tong Weng[a,1], Wenting Lu[a,c], Baihong Sun[a,c], Jiewu Song[a], Qian Zhang[a], Martin Kunz[d], Bihan Wang[e] and Elissaios Stavrou[a,b,c,*]

[a]*Department of Materials Science and Engineering, Guangdong Technion-Israel Institute of Technology, Shantou, 515063, China*

[b]*Guangdong Provincial Key Laboratory of Materials and Technologies for Energy Conversion, Guangdong Technion-Israel Institute of Technology, Shantou, 515063, China*

[c]*Department of Materials Science and Engineering, Technion-Israel Institute of Technology, Haifa, 3200003, Israel*

[d]*Advanced Light Source, Lawrence Berkeley Laboratory, Berkeley, California, 94720, USA*

[e]*Deutsches Elektronen-Synchrotron DESY, Hamburg, 22607, Germany*

∗Corresponding authors: A.S. Ahmad: azkar.ahmad@gtiit.edu.cn; E. Stavrou elissaios.stavrou@gtiit.edu.cn, tel:+86 19880843316.

[1]These authors contributed equally to this work.

**Abstract**

We report the synthesis of the AuGa$_2$ intermetallic compound, using a direct reaction of the relevant elements at room temperature and at very low pressure. The pressure needed to synthesize the AuGa$_2$ compound is below 1 GPa, that is at the lower limit of modern large volume presses, routinely used to synthesize other commercially available materials. This study presents a new method of synthesizing AuGa$_2$, which is much more cost efficient and environmentally friendly than the previously used high-temperature synthesis techniques, and will open new possibilities of synthesizing other intermetallic compounds using high-pressure athermal techniques.

## 1. Introduction

The intermetallic compounds are of fundamental interests because of their technological applications in industrial growth. Like other intermetallic compounds, the compounds belonging to the AuX$_2$ (X=In, Ga, Al) family, are used in a numerous application, ranging from jewelry (colored gold) to electronic interfaces and brazing materials. Specifically, AuGa$_2$, is used in jewelry as "blue gold", in electronic applications as solder joints (between GaAs and gold contacts), and also, for lead (that is highly toxic) free brazing. AuGa$_2$ at ambient conditions AuGa$_2$ adopts a cubic fluorite-type (CaF$_2$-type) crystal structure [1], with a lattice constant of ≈6.076 Å. In this structure, the Au atoms form a face-centered-cubic (fcc) lattice (4a Wyckoff positions) and Ga atoms lie at the 8c Wyckoff position (±[1/4, 1/4, 1/4]).

Although AuGa$_2$ is both technologically and scientifically important, the methods used previously for its synthesis makes the corresponding cost extremely high and the synthesis environmentally unfriendly. To our knowledge, in the past, the only way to synthesize AuGa$_2$, was through a high-temperature reaction between the relevant elements in an argon atmosphere using electrical arc melting [2] or high-temperature furnaces [3, 4]. The high temperatures needed (≈ 1000 $^0$C) and the need for Argon atmosphere (to avoid oxidation) make the above techniques very cost ineffective and also environmentally unfriendly. Recently, pressure has been used for synthesizing novel intermetallic compounds [5], however, it has not been exploited for the synthesis of AuX$_2$ compounds. Here we present a novel high-pressure method to synthesize the $AuGa_2$ compound at room temperature (RT) as an alternative technique to the high-temperature arc melting and furnace methods. Our new method is relatively straightforward, cost-effective and environmentally friendly.

## 2. Materials and Methods

Two independent high-pressure synthesis methods have been used for the synthesis of the $AuGa_2$ compound. For both methods, diamond anvil cells (DAC) with diamond culets of 500 $\mu$m diameter were used. The specimen chamber was constructed using a pre-indented Rhenium gasket with a thickness of 50 $\mu$m and a central hole diameter of ≈ 200 $\mu$m. In-situ high pressure angle-dispersive synchrotron X-ray diffraction (XRD) measurements were performed at the Advanced Light Source, Lawrence Berkeley National Laboratory (Beamline 12.2.2) and at DESY (Beamline P02.2) for synthesis and structural verification. Integration of powder diffraction patterns was performed using the DIOPTAS program [6]. Calculated XRD patterns were produced using the POWDER CELL program [7] assuming Debye rings of uniform intensity. XRD patterns indexing has been performed using the DICVOL program [8]. For both synthesis methods, high-purity Au (99.95%, Alpha Aesar) was used that was mixed with an excess of either high-purity Ga (99.99%, Aladdin Scientific) or high-purity (99.999%) eutectic GaInSn alloy (68.5/21.5/10% of Ga, In and Sn, respectively). In both methods, to avoid oxidation of the Ga or the GaInSn alloy, the loadings were performed inside a glovebox under an argon atmosphere. Au was also used as the pressure marker, through a gold Equation of state (EOS) [9].

## 3. Results and discussion

Fig. 1 shows the high-pressure in-situ XRD pattern collected at 0.1 GPa, the minimum pressure achieved (*i.e.* the DAC was initially closed) in this study. A clear indication of the synthesis of a new compound was observed, based on observation of intense Bragg peaks that cannot be explained by neither the FCC Au crystal structure nor the known structures of Ga. An agreement is observed between the XRD pattern of the experimentally synthesized compound in this study and the calculated XRD pattern of the CaF$_2$-type structure of AuGa$_2$, according to previous structural studies at ambient conditions [1]. This documents the successful synthesis of the AuGa$_2$ compound, and the direct reaction between the Au powder and the Ga liquid even at the minimum achieved pressure of this study.

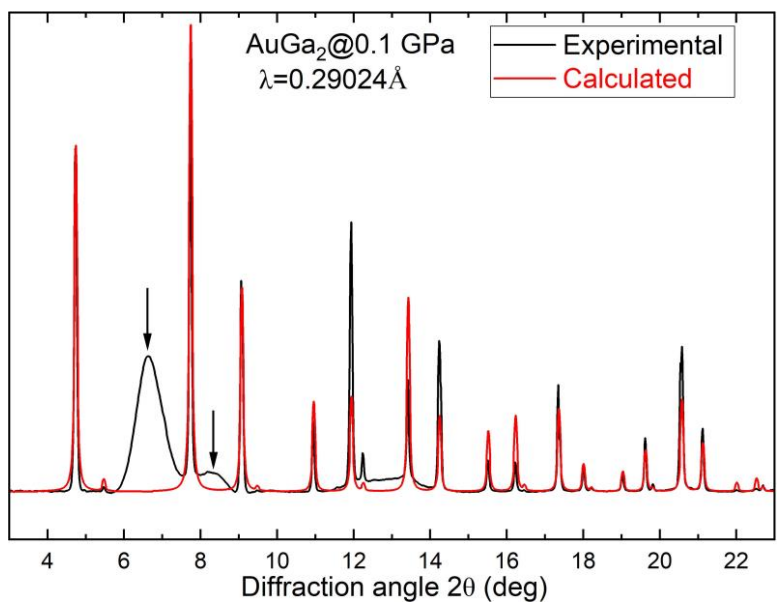

Figure 1: X-ray diffraction patterns of the experimentally synthesized AuGa$_2$ in this study in comparison to the calculated pattern for the CaF$_2$-type structure of AuGa$_2$. The broad peaks indicated by black inverted arrows originate from underacted excess of liquid Ga. The X-ray wavelength is $\lambda$=0.2904 Å.

Given that the reaction was carried out in an excess amount of Ga, some of the Gallium remained unreacted. It should be stressed that in the current high-pressure synthesis method, the pressure needed for the synthesis of AuGa$_2$ compound is below 1 GPa, that is at the lower limit of modern large volume presses or even hydrothermal 'bombs' and therefore, the current high-pressure synthesis method is technically feasible. Importantly, the current method offers a much economical and environmentally friendly method compared to previously reported synthesis methods. After the synthesis of AuGa$_2$ was confirmed, we further studied its structural evolution under pressure. Interestingly, in agreement with the previous reports [10, 11], all the XRD patterns could be indexed with the CaF$_2$-type structure, indicating the absence of any pressure-induced phase transition up to the highest pressure, 7 GPa, examined in the current study. Above 8 GPa all Au was consumed, *i.e.* reacted with Ga towards formation of the AuGa$_2$ compound, even at the places that initially Au was in excess, and thus pressure determination was not possible. From the relevant Le Bail refinements, the cell volumes of AuGa$_2$ were deduced and are plotted in Fig. 2, and compared with previous high-pressure studies of AuGa$_2$.

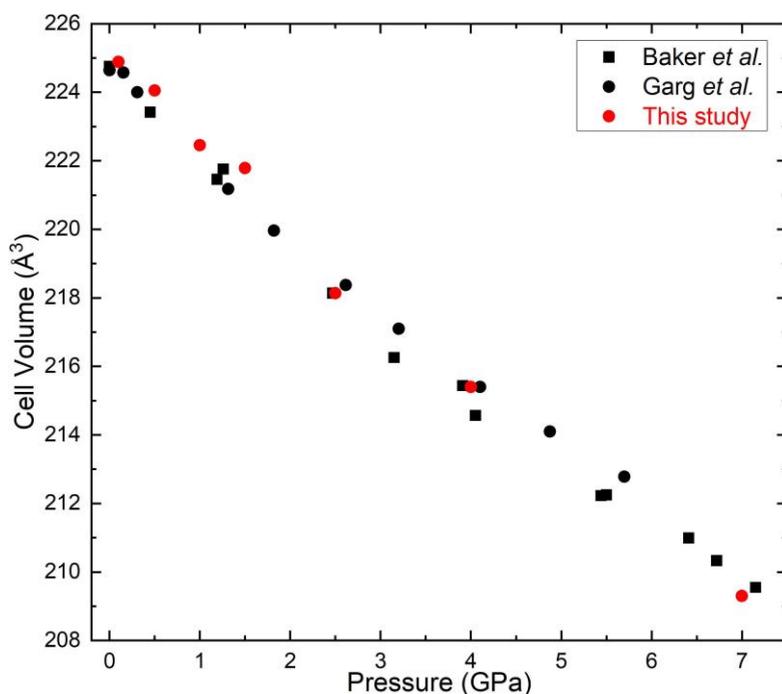

Figure 2: Pressure-dependence of the unit cell volume of the CaF$_2$-type structure of AuGa$_2$ according to this study (red circles). The results from Ref. [11] and Ref. [10] are also plotted for comparison.

The agreement between all sets of EOSs data further confirms the successful synthesis of the AuGa$_2$ compound in this study. Despite the non-hydrostatic conditions in our study, the AuGa$_2$ cell volume at each pressure of this study is similar with the previous high-pressure structural studies using pressure transmitting media (PTM). Most importantly, no pressure-induced phase transition was observed. Thus, CaF$_2$-type structure of AuGa$_2$ is nearly insensitive to the hydrostaticity level, at least in the case of the relatively low bulk modulus of Ga, B$_0$=40-60 GPa [12]. With decreasing pressure, the

AuGa$_2$ compound remained stable after full release to atmospheric pressure.

We also explored the possibility of the synthesis of AuGa2 using an eutectic alloy of gallium (Ga:68.5%), indium (In:21.5 %), and tin (Sn:10.0%) [13]. An identical AuGa2 compound was formed, as documented by XRD measurements, see Fig. 3. In this run, the AuGa2 compound was observed already at 0.3 GPa. This documents the selectivity of the Au+Ga reaction, over the Au+In one that might results to AuIn2, in the liquid state of the eutectic alloy. Indeed, no indication of the formation of the AuIn2 compound (that will have a higher lattice parameter $a$=6.502Å) was observed. This confirms that the synthesis of AuGa2 is highly favorable even at low pressure.

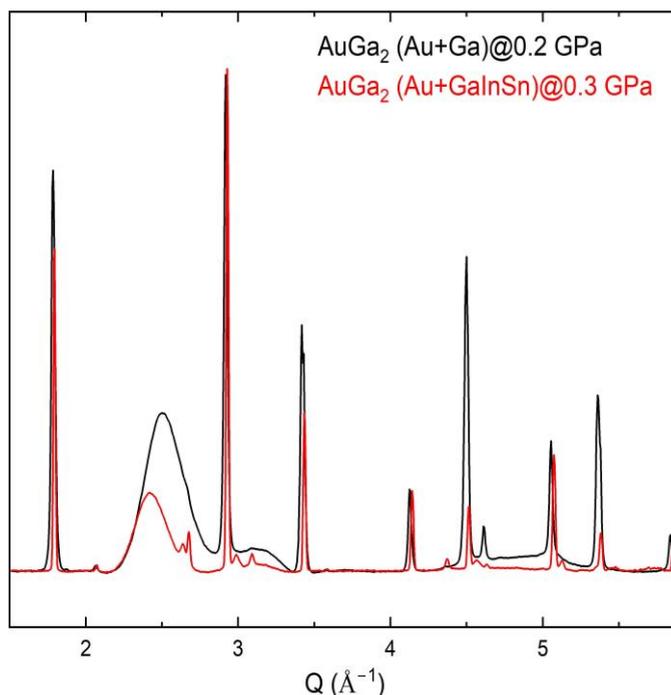

Figure 3: Comparison of the XRD patterns of AuGa$_2$ synthesized under pressure starting from Au+Ga mixture (black) and Au+GaInSn eutectic alloy (red).

4. **Summary**

We have synthesized the AuGa$_2$ compound, using a direct reaction of a mixture of the relevant elements at room temperature and very low pressure (*i.e.* 0.1 GPa) inside a DAC. This high-pressure synthesis of AuGa$_2$ compound is the first ever report on high-pressure synthesis of the AuX$_2$ compounds' family. After the synthesis, the AuGa$_2$ compound was further compressed up to ≈ 8 GPa and its structure was found to be stable. The same compound was also synthesized starting from a Au+GaInSn eutectic alloy mixture. Our study, offers a unique possibility of synthesizing other intermetallic compounds using high-pressure synthesis route using DACs or large volume presses and bypassing the costly and environmentally unfriendly high-temperature techniques.

## Acknowledgments

The financial support from Guangdong Technion-Israel Institute of Technology (Grant No. ST2100002) and MATEC (Grant No. 2022B1212010007, Guangdong Department of Science and Technology) is acknowledged. Beamline 12.2.2 at the Advanced Light Source is a DOE Office of Science User Facility under contract no. DE-AC02- 05CH11231. We also acknowledge DESY (Hamburg, Germany), a member of the Helmholtz Association HGF, for the provision of experimental facilities. Parts of this research were carried out at PETRA III beamline P02.2.

## Conflict of Interest

The authors have no conflicts to disclose.

## Data availability

The data that support the findings of this study are available from the corresponding author upon reasonable request.